\documentclass[usenatbib]{mn2e}
\pdfoutput=1
  \usepackage[T1]{fontenc}
  \usepackage{times}
  \usepackage[pdftex]{graphicx,color}
  \usepackage{subfigure}
  \usepackage{afterpage}
\title[A multi-pixel beamformer using an interferometric array]
{A multi-pixel beamformer using an interferometric array and its application towards localisations of newly discovered pulsars}
\author[Jayanta Roy et al.]
{Jayanta Roy$^1$, Bhaswati Bhattacharyya$^2$ \& Yashwant Gupta$^1$ \\\\
$^1$National Centre for Radio Astrophysics, TIFR, Pune University Campus, Post Bag 3,
Pune 411 007, India\\
$^2$Inter-university Centre for Astronomy and Astrophysics, Pune University Campus, Post Bag 4, Pune 411 007, India}

\date{Accepted. Received}
\begin{document}
\label{firstpage}
\maketitle
\pagerange{\pageref{firstpage}--\pageref{lastpage}} \pubyear{2012}
\def\LaTeX{L\kern-.36em\raise.3ex\hbox{a}\kern-.15em
    T\kern-.1667em\lower.7ex\hbox{E}\kern-.125emX}

\begin{abstract}
We have developed a multi-pixel beamformer technique, which can be used for
enhancing the capabilities for studying pulsars using an interferometric array.
Using the Giant Metrewave Radio Telescope (GMRT), we illustrate the application 
of this efficient technique, which combines the enhanced sensitivity of a 
coherent array beamformer with the wide field-of-view seen by an incoherent array beamformer. 
Multi-pixel beamformer algorithm is implemented using the recorded base-band data. With 
the optimisations in multi-pixelisation described in this paper, it is now possible to 
form 16 directed beams in real-time. 
We discuss a special application of this technique, where we use continuum imaging followed by the
multi-pixel beamformer to obtain the precise locations of newly discovered
millisecond pulsars with the GMRT.
Accurate positions measured with single observations enable highly sensitive follow-up studies 
using coherent array beamformer and rapid follow up at higher radio frequencies and other wavelengths. 
Normally, such accurate positions can only be obtained from a long-term pulsar timing program. 
The multi-pixel beamformer technique can also be 
used for highly sensitive targeted pulsar searches in extended supernova remnants. In addition this 
method can provide optimal performance for the large scale pulsar surveys using multi-element 
arrays.    
\end{abstract}

\label{firstpage} \pagerange{\pageref{firstpage}--\pageref{lastpage}} %
\pubyear{2012}


\section{Introduction}                \label{sec:intro}    
A radio interferometric array is primarily designed to map the sky brightness distribution by
measuring the instantaneous cross-correlations (visibilities) for all possible baselines 
connecting individual antenna pairs. Each baseline vector changes with time due to the rotation of the Earth, 
providing new sets of visibility measurements, and thereby implementing a multi-element Earth
rotation aperture synthesis telescope \citep{Ryle46}. In addition to the interferometric imaging mode, such a 
multi-element radio telescope array is required to be configured to work as a single dish
(called ``beamformer'') for studying compact objects like pulsars, which are effectively point 
sources even for the largest baseline of the array, but demand higher time resolution observations. 
With a flexible back-end design, an interferometric array provides the opportunity of combining 
the imaging mode with beamforming modes that enable new types of pulsar studies.

The GMRT is a multi-element aperture synthesis telescope consisting of 30 antennas, each of 
45 m diameter, spread over a region of 25 km diameter and operating at 5 different wave bands from 
150 MHz to 1450 MHz \citep{Swarup97}. The dual polarised voltage signals from individual antennas are 
eventually converted to base-band signals and are fed to the digital signal processing back-end. 
The recently developed GMRT Software Back-end (GSB), built using a Linux cluster of 48 Intel 
Xeon servers, is a fully real-time back-end for up to 32 dual polarized antenna signals, Nyquist 
sampled at 33 or 66 MHz \citep{Roy10}. The GSB supports an FX correlator (where the cross-multiplication 
(X) is done after the frequency analysis (F), \cite {Thompson94}) and a beamformer.   
The GSB beamformer provides two modes of operation: (i) an incoherent array mode, where the voltage 
samples from the selected antennas are added after converting to intensities, (ii) a single-pixel coherent array 
mode, where the voltage signals from the selected antennas are first added coherently then converted 
to intensity samples by squaring \citep{Gupta00}.
Finally the intensity products are integrated to desired time resolution. 

We have developed an efficient technique, the multi-pixel beamformer, which is currently implemented using the recorded 
base-band data. The optimisation techniques employed in multi-pixelisation makes it capable of 
forming 16 directed beams in real-time. 
The multi-pixel beamformer combines the enhanced sensitivity of 
a coherent array with the wide field-of-view seen by an incoherent array.
Pulsars originally discovered in surveys with the incoherent array have large uncertainties 
in positions because of the large primary beam, which prevents highly-sensitive follow-ups of those pulsars 
using the coherent array or at higher frequencies with narrower primary beams. The multi-pixel beamformer 
can be used to rapidly localise the pulsar, to an accuracy equal to the synthesized beam of the array 
($<$ 10\arcsec), using a single snapshot observation. With traditional single-dish pulsar observations, 
such precise localisations can only be achieved from long-term pulsar timing programs. We have successfully 
demonstrated this efficient technique by determining accurate positions of two newly discovered Fermi 
millisecond pulsars (MSPs) using the GMRT. 


\section{Single-pixel beamformer and base-band recorder at the GMRT}  \label{sec:GSB} 
The GMRT correlator utilises an FX approach as described in \cite{Roy10}. The integral parts of the 
antenna-dependent geometrical path delays are compensated prior to spectral decomposition (FFT), whereas 
the fractional parts are combined with fringe derotation and applied at the post-FFT stage. 
In parallel with the multiply-accumulate operations, the spectral voltages, after correcting for 
antenna-based gain offsets and time delays, are fed to an array combiner to generate the incoherent 
and coherent beamformer output. In addition to the gain calibration the antenna-based phase offsets are 
also calibrated out for the coherent beamformer mode. The visibilities obtained from a calibrator 
source are used to solve for both the broadband and narrowband phase offsets, which are then applied 
at the post-FFT stage along with the fringe correction. Such interleaved calibrator observations in 
every 2 h are required to optimise the coherent array sensitivity at lower frequencies (e.g. 325 MHz).
The GSB produces an incoherent beam and a single-pixel coherent beam (formed for the pointing centre of 
the array) at 30$\mu$s time resolution with 512 spectral channels, which translates to 32 MB/s output 
rate for total intensity products (after adding two polarised intensities). Then the incoherent 
and coherent array outputs have to be corrected for the frequency-dependent delays due to the dispersion 
in the interstellar medium for the target pulsar. 

In addition to the correlator and the single-pixel beamformer, the GSB cluster also supports base-band recording, 
which is direct streaming of raw Nyquist samples from the antennas into the array of storage disks. The 
aggregate streaming rate is 3.3 TB/h. The raw voltage samples are piped to an off-line cluster, where 
the multi-pixel beamformer algorithms are implemented.
\begin{figure}
\begin{center}
\includegraphics[angle=270,width=0.5\textwidth]{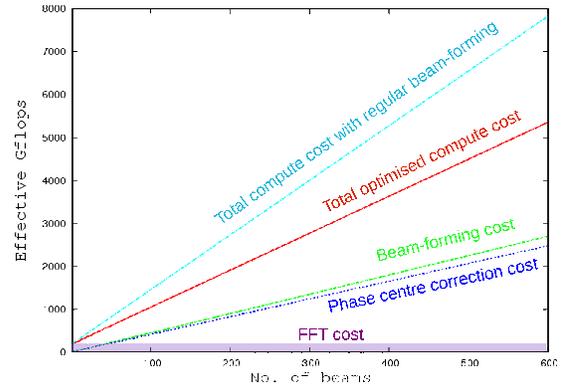}
\caption[Compute cost for multi-pixel beamformer]{Compute cost for the multi-pixel beamformer as a function
of the number of beams. Approximately 600 beams are required to cover the full field-of-view
using around 50\% of the GMRT array. We achieve a $>30$\% reduction in the compute cost for multi-pixelisation
while uniformly covering large fraction of the field-of-view. This can be seen from the solid red-line
(plotting total optimised compute cost) and dash-dot sky-blue-line (plotting the total compute cost using
regular brute-force beam-forming).}
\label{multibeam_cost}
\end{center}
\end{figure}

\section {Multi-pixel beamformer}  \label{sec:Multi-pixel beamformer}
We have achieved the multi-pixelisation of the field-of-view
by forming multiple beams steering the phase centre. For the GMRT specifications, the arithmetic computational cost for 
the coherent beamformer is $\sim$ 62 times lower than that of the cross-multiplier-adder \citep{Roy10}. However, including the 
non-arithmetic overheads due to the input/output memory operations, the coherent beamformer is $1.8\times$ cheaper than the cross-multiplier-adder. 
Thus we preferred a multi-pixel beamformer over combining 
the auto and cross visibilities with different phases. \cite{Roy10} 
estimated the compute cost for a single beamformer at around 193 Gflops, with a break up of 181 Gflops for 
the FFT, 8.25 Gflops for the antenna-based phase centre correction, and 4.5 Gflops for the coherent beamforming 
operation. We employ multi-pixelisation at the post-FFT stage, thus the phase centre correction and the 
beamforming cost scale with the number of beams, whereas the FFT cost remains fixed, as seen in Fig. \ref{multibeam_cost}.

For a 325 MHz operating frequency, the pixelisation can be done efficiently up to an 
angular scale of 1\arcmin, which uses around 72\% of the GMRT array. The perturbations in the ionospheric phases
(which are severe at lower frequencies) limits the baseline length over which the array can be coherently added 
with optimal efficiency. Thus 6400 coherent beams are required for multi-pixelisation (with 1\arcmin\ resolution) 
of full field-of-view (80\arcmin\ at 325 MHz) with a sensitivity improvement $\sim 5\times$ with respect to 
the incoherent array. The required number of beams is independent of the operating frequency, since the 
pixel angular size and the 
field-of-view scale similarly with frequency. The array configuration of the GMRT allows us to 
employ further optimisations in this technique, as described below. 

The maximum geometrical delay for the GMRT array is 128$\mu$s, corresponding to the longest
baselines of $\sim$ 30 km. Hence the maximum residual 
delay at the edge of the field-of-view is lower than the Nyquist sampling resolution and we can retain 
the integral delay calculated for the pointing centre across the field-of-view. The phase center 
correction cost (including the fractional delay compensation) is optimised by grouping the GMRT array. 
For example, at 325 MHz considering compact core of the GMRT array (which consists of 6 antennas) we 
apply the same phase correction up to 10\arcmin\ from the pointing centre (in order to maintain the 
phase error at less than a degree). 
In addition, a trade-off between the search sensitivity and computational load can be
done by generating relatively fat beams of 3\arcmin\ in size (i.e. $\sim$ 600 such beams will cover
the full field-of-view) using around 50\% of the GMRT array. This provides a sensitivity improvement
of 3$\times$ with respect to the incoherent array, whereas for 72\% of the GMRT array with coherent beams
of 1\arcmin\ in size, the sensitivity improvement is $5\times$. This compromise in sensitivity improvement
allows a reduction in the total compute cost by an order of magnitude. 
In view of the available compute power, our current multi-pixel beamformer is designed to 
generate such fat beams of 3\arcmin\ in size.
Fig. \ref{multibeam_cost} illustrates  the scaling of compute cost of each 
module (such as phase centre correction, beam-forming) with the number of beams. The effect of 
the optmisation in phase centre correction is shown by the growing difference between the solid red-line (total 
optimised compute cost) and dash-dot sky-blue line (the total compute cost using 
regular brute-force beam-forming). The total compute cost for multi-pixelisation is reduced by $>30$\%, while 
uniformly covering large fraction of the field of view. 

The multi-pixel beamformer is currently designed to produce coherently added intensity beams (2 bytes/sample) at 30 $\mu$s time resolution 
with 512 spectral channels using the recorded base-band data. This implies 512 MB/s of output streaming rate for 16 beams, 
which is supported by the quad gigabit networks. Hence with employed optimisation and with available network bandwidth, 
we are currently capable of forming and streaming 16 directed beams in real-time.
Our off-line analysis pipeline uses full floating point version 
of the real-time code (real-time code uses fixed point arithmetic to extract the full vectorized power of the CPU) and 
it is benchmarked to produce $1.5\times$ lower flops than the real-time; i.e. for 16 beams the off-line pipeline takes $1.5\times$  
of the observing time.
In addition to the sensitivity gain from multi-pixelisation, we expect to get further 
improvements in detection significance of a pulsed signal from the fact that the 
incoherent array, being a sum of individual total power detectors, is more vulnerable to the instrumental 
gain fluctuations and the terrestrial radio frequency interference (RFI). However, for the coherent array 
these effects are reduced and the array provides some built-in immunity to RFI 
as the processing pipeline adjusts the antenna phases to correct for the effect of rotation of the sky signals, 
which in turn de-correlates the terrestrial signals. 
\section {Localisation of pulsars using multi-pixel beamformer} \label{sec:localisation}
The optimised multi-pixel beamforming technique described above can be very useful
for large scale blind pulsar surveys, to achieve coherent array sensitivity with
telescope observing time commensurate with an incoherent array coverage of the
survey area. Possible future applications of this are discussed in  \S \ref{sec:summary}.
However, in parallel with multi-pixel beamforming, an interferometric array like the GMRT also allows 
simultaneous imaging of the field-of-view, which provides \textit{a priori} information about the possible locations of interest. 
Continuum imaging followed by multi-pixel beamforming leads to a drastic reduction in the compute cost, 
as we do not need to uniformly cover the field-of-view when candidate source locations can be provided from the image. 
We illustrate the application of multi-pixel beamformer to obtain precise locations of the Fermi MSPs recently discovered at the GMRT.

With the GMRT, as a part of Pulsar Search Consortium, recently we have discovered six millisecond pulsars \citep{Ray12} 
in a 325 and 610 MHz survey of the error boxes of \textit{Fermi} LAT gamma-ray sources. The error boxes of the LAT 
sources can be as large as 18\arcmin, which are conveniently covered by the wide incoherent array beam of the GMRT 
(e.g. the beam-width is 80\arcmin\ at 325 MHz and 40\arcmin\ at 610 MHz). The large solid angle of the incoherent beams also provide the 
possibility of discovering in-beam pulsars at large offsets from the pointing centre that are unrelated to the target LAT source. 
A few of the \textit{Fermi} MSP discoveries have been found to be serendipitous and are not associated with 
the corresponding gamma-ray sources. 

\begin{figure*}
\begin{center}
\includegraphics[angle=270,width=0.7\textwidth]{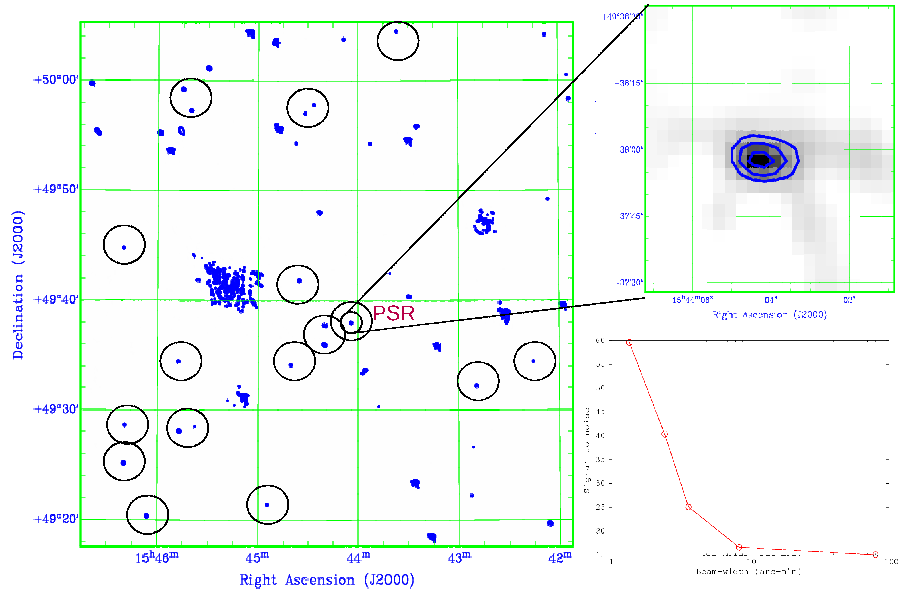}
\caption[Localising PSR J1544+4937]{Figure for localisation of the Fermi MSP J1544$+$4937 (discovered by us with the GMRT)
using continuum imaging followed by a multi-pixel search. The central plot shows the field of PSR J1544$+$4937, with 16 candidate
point sources (circled) on which coherent beams are formed. One of these beams has the pulsar at the center and 
has maximum signal to noise in the search output. The MSP is found at 15$^\mathrm{h}$44$^\mathrm{m}$04\fs166, $+49$\degr37\arcmin57\farcs45. 
The top right panel presents a zoomed image of 1\arcmin region around the MSP.  The bottom right panel illustrates 
the sensitivity improvement with the number of antennas added in coherent array (i.e. when reducing the beam-width).}
\label{J1544+4937}
\end{center}
 \end{figure*}

PSR J1544$+$49 is one of the MSPs we discovered having a period of 2.16 ms and dispersion measure (DM) of 23.2 pc cm$^{-3}$. 
From our discovery observations, the estimate of the mean flux of this pulsar is around 2.5 mJy at 610 MHz.
Considering the beam size at 610 MHz, the position of this MSP is determined to be within $\pm$20\arcmin\ of the pointing centre. 
Since the pulsar is at high Galactic latitude (i.e. we are not limited by sky background temperature) and expected 
to have a steep radio spectrum, we chose 325 MHz as the frequency for follow-up observations. For the imaging and 
multiple pixel beamforming, we recorded base-band data for the nominal target position for a duration of one hour.  
The central 40\arcmin\ region of the field-of-view was imaged. The flagging of erroneous visibilities, e.g. those affected 
by RFI and the calibration for the complex gain are handled by flagcal pipeline 
\citep{Prasad11}. The calibrated visibilities are imaged and deconvolved using standard imaging package AIPS.
To select potential 
pulsar candidate locations, we identified the sources in the image with similar flux values to the mean flux of the MSP 
($\sim$ 10 mJy is estimated with the incoherent array generated from the base-band data). 
Fig. \ref{J1544+4937} illustrates the PSR J1544$+$49 field. We have chosen 16 candidate locations (circled) 
for follow-up with multi-pixel beamforming. 
The coherent beams (3\arcmin\ in size using 50\% of the GMRT array) are formed simultaneously for all the candidate 
sources with the multi-pixel beamformer pipeline (described in \S \ref{sec:Multi-pixel beamformer}) using 
the recorded base-band data. Such fat beams can span more than one point source of interest within a single beam.
The optimisation technique used for reducing the phase correction cost is also effective while forming beams 
near the pointing centre (within a $\sim10$\arcmin\ region), which is applicable for most of the Fermi MSPs.
The multi-pixel beamformer outputs are fed to a search pipeline, which is designed to 
search for periodicity and acceleration in parallel on all these beams. In this search we have used the DM 
of this pulsar obtained from the discovery search analysis. The beam with pulsar at the center gives maximum 
signal to noise. The MSP is found at 15$^\mathrm{h}$44$^\mathrm{m}$04\fs166, +49\degr37\arcmin57\farcs45. 
The zoomed image of a 1\arcmin\ region around the position of the MSP is 
shown in the top right panel. The three contours represent 80\%, 60\% and 50\% of the peak 
flux. The pulsar is detected with 20$\sigma$ significance 
in the continuum image at an offset of 4\farcm3 from the search pointing centre. The sensitivity 
improvement with the number of antennas added in the coherent array 
(i.e. with reducing the beam-width) validates the correctness of the candidate pulsar position, shown in 
the bottom right panel of Fig. \ref{J1544+4937}. The initial incoherent array uncertainty of 80\arcmin\  
first reduces to 8\arcmin\ using the coherent array consisting of the innermost core of the GMRT array. 
Eventually, the sensitivity becomes 4 times that of the incoherent array with the beam-width being 1\farcm4. 
The final positional uncertainty is decided by the synthesized beam used for the continuum imaging.
Using this technique we have determined the position for PSR J1544$+$4937 with an accuracy 5\arcsec, 
which is half of the synthesized beam used in the imaging.

\begin{figure*}
\begin{center}
\includegraphics[angle=270,width=0.7\textwidth]{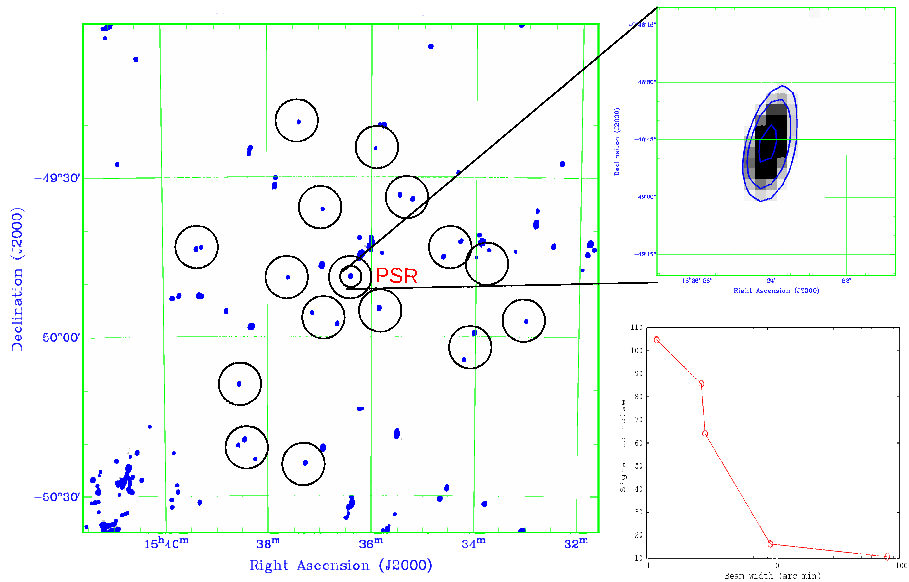} 
\caption[Localising PSR J1536-4948]{Same as Fig. \ref{J1544+4937} for Fermi MSP J1536$-$4948 (discovered by us with the GMRT).}
\label{J1536-4948}
\end{center}
\end{figure*}

This method is also successfully applied for localising another MSP, J1536$-$49, which has a period of 3.08 ms 
and a DM of 38.0 pc cm$^{-3}$. From the discovery observations, we estimate the mean flux of this pulsar to be 
around 12 mJy at 325 MHz.  We recorded one hour of base-band data for this MSP.
Fig. \ref{J1536-4948} illustrates the 80\arcmin\ field for this pulsar. Coherent beams 
were formed on 16 chosen candidates (circled) having estimated fluxes similar to the pulsar flux. 
The MSP was found at 15$^\mathrm{h}$36$^\mathrm{m}$24\fs016, $-49$\degr 48\arcmin45\farcs39. The zoomed image 
of 1\arcmin\ region around this MSP is shown in the top right panel. The three contours represent 
90\%, 60\% and 50\% of the peak flux. The pulsar is detected with 11 $\sigma$ significance in 
the continuum image at an offset of 1\farcm8 from the search pointing centre. Similar to PSR J1544$+$4937, 
the sensitivity improvement (up to 10 times) with the reduction in the beam-width 
(illustrated in the bottom right panel of Fig. \ref{J1536-4948}), shows that the pulsar position is 
correctly estimated. In addition to the sensitivity gain from multi-pixelisation, we obtained further 
improvement of the detection significance for this extreme southern pulsar due to fact that the incoherent 
array being a sum of individual total power detectors with large beam-width is more affected by the terrestrial RFI coming from
near the horizon. We have determined the accurate position for PSR J1536$-$4948 with an accuracy 14\arcsec. 
\section{Summary and future scope}  \label{sec:summary}
We have developed a multi-pixel beamforming technique, which can be used for enhancing the capabilities of studying pulsars using 
an interferometric array. As an example of the application of this technique, we have used 
continuum imaging followed by multi-pixel beamforming to determine the 
accurate positions of two MSPs, which we have discovered at the GMRT in searches of \textit{Fermi} LAT 
unassociated gamma-ray sources as part of the \textit{Fermi} Pulsar Search Consortium. With a single 
snapshot observation this technique enables us to significantly reduce the large positional uncertainties 
associated with incoherent array discovery (e.g. from 80\arcmin\ or 40\arcmin\ to $\sim$ 10\arcsec). 
With single dish telescopes, such precise position determinations can be achieved with long-term pulsar 
timing program. The initial position of a newly-discovered pulsar can also be improved up to certain extent using 
a grid of observations at higher frequency. This costs substantial telescope time and can be hindered by 
scintillation and the steep spectra of pulsars. Prompt knowledge of accurate positions allows us to make highly 
sensitive follow-up observations of these pulsars using single-pixel coherent beam investing much less telescope time.
In addition more efficient timing for the pulsar can be carried out by reducing two degrees 
of freedom in parameter space \citep{Lorimer04} for getting the initial solution. This is very helpful in timing the MSPs in 
tight binaries with large degrees of freedom. Knowledge of the accurate pulsar positions allows for rapid 
follow-up at higher radio frequencies at different telescopes. Such arcsecond localization also facilitates the study of 
X-ray and optical counterparts for these MSPs. In addition, large positional uncertainties associated with the discoveries 
of pulsars from low frequency wide field surveys will get benefited from the multi-pixelisation of the field of view.
   
Searching for young pulsars in the extended Galactic Supernova Remnants (SNRs) can benefit from 
the multi-pixel beamforming technique.
The conventional single-pixel search with narrow field-of-view can miss a young pulsar 
if the positional uncertainty within the remnant is larger than the beam width. 
Pulsars in SNRs with large birth velocities may travel significant distances from the remnant centroids 
\citep{Lyne94} and are detected at the peripheries of SNRs (\cite{Caraveo93}, \cite{Lorimer98}), which creates a selection bias. 
Thus simultaneous multi-pixel search with reasonably wide field-of-view is necessary for such 
candidate SNRs. For example, in case of a filled-centre SNR with an evidence of centrally bright emission 
region of size $\sim$ 10\arcmin, about 40 beams (each with 1.5\arcmin\ in size) are needed to be synthesized at 610 MHz. 
This gives a sensitivity gain of a factor of 3 with respect to the incoherent array, resulting in a saving in 
observing time at least by a factor of 9, which is very important for reducing the required telescope time 
for a multi-purpose telescope like the GMRT.     

The simultaneous highly sensitive multi-pixel search uniformly covering the full field-of-view can be a 
very useful technique for getting optimum performance from blind pulsar surveys with multi-element 
radio telescope arrays of the future. This technique will be particularly essential in finding 
weaker pulsars below the incoherent array detection threshold. Moreover this multi-pixel coherent 
search can overcome the sensitivity degradation seen by the incoherent array due to instrumental 
gain fluctuations and variable RFI environments. 
Our current studies using the GMRT can work as a test-bed for the 
future developments of bigger arrays. With the aid of further optimisation, GPU (graphical processing unit) 
based processing and the availability of 10G base-T or infiniband networks, our current design holds promises 
for large scale pulsar surveys using interferometric arrays.   
\section{Acknowledgments}
We thank the computer group at GMRT and NCRA for helping in setting up the compute cluster for multi-pixelisation. 
Development of multi-pixel beamformer was motivated by discoveries of Fermi MSPs from our GMRT survey. 
Dr. David Thompson and Dr. Paul Ray had helped us to join in the Pulsar Search Consortium. We thank Dr. Paul Ray 
for providing detailed comments on the manuscript. We thank Prof. Dipankar Bhattacharya for being an active member 
of GMRT team and for his critical comments. We would like to thank Prof. Ue-Li Pen for insightful discussions on the 
optimisation of the multi-pixel beamforming algorithm. 
We acknowledge the support of telescope operators during our data intensive base-band recording observations. The GMRT is run by the National 
Centre for Radio Astrophysics of the Tata Institute of Fundamental Research.

\end{document}